\pdfoutput=1
\documentclass{article}
\usepackage{spconf,amsmath,graphicx,hyperref}
\usepackage{cite}
\usepackage{amsmath,amssymb,amsfonts}
\usepackage{algorithmic}
\usepackage{graphicx}
\usepackage{xcolor}
\usepackage{multirow,multicol}
\usepackage{booktabs, siunitx}
\usepackage{setspace}

\usepackage{makecell}
\usepackage{color, colortbl}
\definecolor{Gray}{gray}{0.9}
\usepackage{subcaption}
\usepackage{microtype}
\hypersetup{hidelinks}

\title{Deep Filter Estimation from Inter-Frame Correlations for Monaural Speech Dereverberation}
\name{Ui-Hyeop Shin$^{1}$, Jun Hyung Kim$^{1}$, Jangyeon Kim$^{2}$, Wooseok Kim$^{2}$, Hyung-Min Park$^{1,2}$
 \thanks{This work was partly supported by Institute of Information \& communications Technology Planning \& Evaluation (IITP) grant funded by the Korea government (MSIT) (RS-2022-II220989, Development of Artificial Intelligence Technology for Multi-speaker Dialog Modeling) and National Research Foundation of Korea (NRF) grant funded by the Korea government (MSIT) (RS-2026-25470024, Research on Spatial Audio Reasoning with Large Language Models across Diverse Microphone Arrays).}}

\address{
    $^1$Department of Electronic Engineering, $^2$Department of Artificial Intelligence\\
    Sogang University, Seoul, Republic of Korea\\
    {\normalsize \{dmlguq123, imalbert, jykim97, wooseokkim, hpark\}@sogang.ac.kr}}

\begin{document}
\ninept
\maketitle
\begin{abstract}
Speech dereverberation with a distant microphone is challenging because reverberation is correlated with the target speech, and models trained on simulated data often generalize poorly to real recordings. 
We propose IF-CorrNet, a correlation-to-filter architecture for monaural dereverberation. 
Instead of feeding raw complex STFT coefficients to the network, IF-CorrNet computes inter-frame correlations among neighboring frames at each time-frequency bin and estimates multi-frame deep filters from these features with a dual-path Transformer backbone. 
This design makes inter-frame dependencies explicit at the network input while retaining a multi-frame filtering output, a pairing motivated by the normal equation of linear multi-frame filtering. 
On the REVERB Challenge corpus, IF-CorrNet achieves the best
CD, LLR, SNR$_{\mathrm{fw}}$, and PESQ among the compared
dereverberation baselines on SimData, and the highest SRMR among
the compared systems on RealData.
The ablation shows higher RealData SRMR with correlation inputs
for both filtering and masking, with filtering adding a further gain.
\end{abstract}
\section{Introduction}
\label{sec:intro}

When speech is captured by a distant microphone, reverberation in indoor environments adds long-tail reflections to the direct speech signal, substantially degrading human intelligibility. These reflections smear temporal and spectral details, leading to reduced clarity and increased error rates in downstream tasks. Unlike additive noise, which is usually uncorrelated with the target speech and thus easier to suppress, reverberation remains strongly correlated with the original signal, making it challenging to remove without introducing artifacts or distorting the desired signal.

Although several dereverberation algorithms have been presented based on statistical modeling of target speech~\cite{wpe_nakatani10, WPE_2012}, they have shown limited performance, especially with a single microphone. With the advent of deep learning, a wide range of dereverberation methods have been proposed~\cite{Ernst18, WangZQ20}, often by treating dereverberation as part of a broader speech enhancement or extraction problem~\cite{Williamson17, TF_GridNet_TASLP, TF_Locoformer}. Conventional approaches feed complex short-time Fourier transform (STFT) coefficients into a network and perform complex masking~\cite{Williamson17, dccrn}, deep filtering~\cite{deepfiltering, DFN}, or direct complex spectral mapping~\cite{Tan19, WangZQ20}. 
Although these methods have shown notable gains, they typically rely on the network to implicitly disentangle the target speech from noisy-reverberant raw inputs, and their generalization from simulated training data to real recordings remains an important concern.

Two-stage networks have also been proposed: the first stage predicts speech components to estimate inter-frame target and noise covariance matrices, while the second stage estimates multi-frame dereverberation filters~\cite{DCN}, yielding improved performance. This is analogous to neural beamforming in the multi-channel case, which estimates beamforming filters from covariance matrices predicted by a separate network~\cite{Zhang21_ICASSP,Zhang22,Ochiai2023_Tracking}.
Multi-frame Wiener and minimum variance distortionless response (MVDR) filters exploit the inter-frame correlation of speech~\cite{Huang2012_MF}, and prior single-microphone enhancement studies have used learned inter-frame correlation quantities for multi-frame MVDR (MF-MVDR) filtering and compared alternative multi-frame filtering formulations~\cite{Tammen2021_MFMVDR,schroter23_interspeech}.

Recently, TF-CorrNet~\cite{TF_CorrNet} was introduced to exploit spatial information using instantaneous correlations in multi-channel input for joint denoising, dereverberation, and separation. 
While TF-CorrNet uses inter-channel spatial correlations, we target temporal correlations across STFT frames for monaural dereverberation.
To this end, we propose IF-CorrNet\footnote{Demo page: {\urlstyle{same}\url{https://dmlguq456.github.io/IF-CorrNet-demo/}}}, which explicitly leverages inter-frame correlations as the primary network input.
Instead of raw STFT coefficients, IF-CorrNet uses pairwise correlations among neighboring frames at each frequency to provide explicit temporal cues for filter estimation.
Our focus is the pairing of explicit inter-frame correlation inputs
with directly estimated multi-frame filters, rather than filters derived from network-predicted inter-frame statistics~\cite{DCN,Tammen2021_MFMVDR}.
As analyzed in Section~\ref{subsec:corr_filter}, the corresponding
second-order statistics reflect reverberant mixing and govern
linear multi-frame filter estimation.
For the backbone, we employ dual-path modules~\cite{dpt_fsnet,TF_Locoformer}
to integrate these features across time and frequency.
Evaluations on the REVERB Challenge~\cite{REVERB} show that IF-CorrNet achieves the highest SRMR among the compared systems on real recordings (RealData), supporting transfer from simulated training data to this real recording condition. An input/output ablation shows RealData SRMR gains from the correlation inputs for both filtering and masking, and a further gain when they are paired with multi-frame filtering.

\section{IF-CorrNet for Dereverberation}
\vspace{-1mm}

Let $X_{tf}\in\mathbb{C}$ denote the observed STFT coefficient at
time frame $t$ and frequency bin $f$, where $1\leq t\leq T$ and
$1\leq f\leq F$. Using an approximate STFT-domain convolutive model, we write
\vspace{-1mm}
\begin{equation}
X_{tf}
=
H_f S_{tf}
+
\sum_{\tau=0}^{L_0} R_{\tau f} S_{(t-\tau)f}
+
N_{tf},
\label{eq:input}
\end{equation}
where $S_{tf}$ is the source speech, $H_f$ is the direct-path transfer function, and $R_{\tau f}$ represents the convolutive transfer terms associated with early reflections and late reverberation. 
Here, $L_0$ denotes the maximum reverberation lag, and $N_{tf}$ is additive noise. 
Our goal is to recover the desired direct signal $D_{tf}=H_fS_{tf}$ by suppressing the reverberant and noise components. 

\subsection{Inter-frame correlations for deep filter estimation}

The delayed speech components in~\eqref{eq:input} cause neighboring observations to share contributions from the same source frames.
Their inter-frame correlations therefore contain information about the reverberant temporal structure, together with the intrinsic
correlations of speech. Rather than relying solely on learned aggregation of raw STFT coefficients, we explicitly provide
pairwise products between the frames that are subsequently combined by the output filter.

Specifically, let the multi-frame observation vector be $\mathbf{x}_{tf}=[X_{(t-L)f},\dots,X_{tf},\dots,X_{(t+L)f}]^{T}\in\mathbb{C}^{2L+1}$,
where $L$ denotes its half-width and is distinct from the reverberation lag $L_0$.
We compute the inter-frame correlation matrix as
\begin{equation}
\mathbf{Z}^{(0)}_{tf}
=
\mathbf{x}_{tf}\mathbf{x}_{tf}^{H}
\in
\mathbb{C}^{(2L+1)\times(2L+1)}.
\label{eq:raw_corr}
\end{equation}
Its diagonal entries contain frame-wise powers, so magnitude information, which is known to benefit enhancement~\cite{TF_GridNet_TASLP,cmgan}, is naturally included, while its off-diagonal entries encode pairwise magnitude products and relative phases. This is an instantaneous outer product rather
than a time-averaged covariance estimate; we retain the full
matrix despite its Hermitian redundancy.
To compress the dynamic range of these power-scaled features,
we apply generalized phase transform (PHAT-$\beta$) weighting~\cite{TF_CorrNet,PHAT_beta} as $\mathbf{Z}_{tf}=\mathbf{Z}^{(0)}_{tf}\oslash|\mathbf{Z}^{(0)}_{tf}|^{\beta}$,
where $\oslash$ denotes element-wise division, and the magnitude
and power operations are also element-wise.
We set $\beta=0.5$. Writing the $m$th component of
$\mathbf{x}_{tf}$ as $x_m=\rho_m e^{j\theta_m}$ gives $[\mathbf{Z}_{tf}]_{mn}
=\sqrt{\rho_m\rho_n}\,e^{j(\theta_m-\theta_n)}$.
The diagonal entries are therefore magnitudes, while the
off-diagonal entries combine geometric-mean magnitudes with relative
phases.

The network integrates these weighted correlations across time and frequency to estimate a complex filter
$\mathbf{w}_{tf}\in\mathbb{C}^{2L+1}$.
The filter is applied to the corresponding observation vector:
\vspace{-1mm}
\begin{equation}
Y_{tf}
=
\mathbf{w}_{tf}^{T}\mathbf{x}_{tf}.
\label{eq:filtering}
\end{equation}
The observation vector includes future frames, making this formulation
non-causal. Its finite filter support is distinct from the broader
temporal context used by the network to estimate the coefficients.

\subsection{Connection to dereverberation filtering}
\label{subsec:corr_filter}

To relate the correlation features to reverberation, let $M\hspace{-1mm}=\hspace{-.8mm}2L\hspace{-.4mm}+\hspace{-.4mm}1$
and combine the direct and reverberant terms as
$a_{\tau f}\hspace{-1mm}=\hspace{-1mm}H_f\delta_{\tau,0}\hspace{-.4mm}+\hspace{-.4mm}R_{\tau f}$.
The source coefficients contributing to $\mathbf{x}_{tf}$ form
$\mathbf{s}_{tf}=[S_{(t-L-L_0)f},...,S_{(t+L)f}]^T
\in\mathbb{C}^{M+L_0}$.
Within the STFT-domain model in~\eqref{eq:input},
\vspace{-1mm}
\begin{equation}
\mathbf{x}_{tf}=\mathbf{A}_f\mathbf{s}_{tf}+\mathbf{n}_{tf},
\qquad
D_{tf}=\mathbf{d}_f^T\mathbf{s}_{tf},
\label{eq:stacked_reverb}
\end{equation}
where $\mathbf{A}_f\hspace{-.7mm}\in\hspace{-.7mm}\mathbb{C}^{M\times(M+L_0)}$ is the
convolution matrix formed from $a_{\tau f}$, $\mathbf{n}_{tf}$
stacks the noise, and $\mathbf{d}_f$ selects $S_{tf}$
and scales it by $H_f$.
For this analysis, fix the acoustic condition and suppress $(t,f)$.
Define $\mathbf{R}_{ss}\hspace{-.4mm}=\hspace{-.4mm}\mathbb{E}[\mathbf{s}\mathbf{s}^H]$ and
$\mathbf{R}_{nn}\hspace{-.4mm}=\hspace{-.4mm}\mathbb{E}[\mathbf{n}\mathbf{n}^H]$.
Assuming $\mathbb{E}[\mathbf{s}\mathbf{n}^{\hspace{-.4mm}H}]=\mathbf{0}$ gives
\begin{equation}
\begin{aligned}
\mathbf{R}_{xx}
&=\mathbb{E}[\mathbf{Z}^{(0)}]
 =\mathbf{A}_f\mathbf{R}_{ss}\mathbf{A}_f^H+\mathbf{R}_{nn},\\
\mathbf{p}_{xD}
&=\mathbb{E}[\mathbf{x}D^*]
 =\mathbf{A}_f\mathbf{R}_{ss}\mathbf{d}_f^*.
\end{aligned}
\label{eq:reverb_moments}
\end{equation}
Thus, the observation correlations reflect the reverberant transfer
structure together with the source's intrinsic temporal correlations.
No independence between direct and reverberant speech, or between
source frames, is required.

For a fixed candidate filter $\mathbf{w}$, the effective source-response
coefficients are $\mathbf{A}_f^T\mathbf{w}$.
Their mismatch from the direct-path target is
$\mathbf{e}_w\hspace{-.5mm}=\hspace{-.5mm}\mathbf{A}_f^{\hspace{-.5mm}T}\hspace{-.5mm}\mathbf{w}\hspace{-.5mm}-\hspace{-.5mm}\mathbf{d}_f$.
The linear mean-squared error can then be written as
\begin{equation}
\mathbb{E}[|D-\mathbf{w}^T\mathbf{x}|^2]
=\mathbf{e}_w^T\mathbf{R}_{ss}\mathbf{e}_w^*
 +\mathbf{w}^T\mathbf{R}_{nn}\mathbf{w}^*.
\label{eq:filter_mse}
\end{equation}
The two terms measure source-weighted mismatch from the target response
and filtered noise power, respectively.
The minimizing linear filter satisfies
\begin{equation}
\underbrace{\mathbb{E}[\mathbf{Z}^{(0)}]}_{\mathbf{R}_{xx}}
\mathbf{w}_{\mathrm{opt}}^*
=\mathbf{p}_{xD}
=\mathbf{A}_f\mathbf{R}_{ss}\mathbf{d}_f^*.
\label{eq:normal}
\end{equation}
The same correlation matrix therefore governs the coupling among
coefficients used to reshape the reverberant response.
This connects explicit correlation features to multi-frame filter
estimation~\cite{Huang2012_MF,Tammen2021_MFMVDR,schroter23_interspeech}.
Although $\mathbf{Z}^{(0)}_{tf}$ adds no information beyond $\mathbf{x}_{tf}$
itself, it explicitly exposes the pairwise interactions that constitute
$\mathbf{R}_{xx}$, an inductive bias aligned with linear multi-frame
filter estimation.
IF-CorrNet learns input-dependent filters from the PHAT-$\beta$-weighted instantaneous products using $\ell_1$-based supervision on the filtered output.
The ensemble moments above provide an
analytical reference; the network does not explicitly estimate
$\mathbf{A}_f$ or solve~\eqref{eq:normal}.

\begin{figure}[t]
    \centering
    \includegraphics[width=0.47\textwidth]{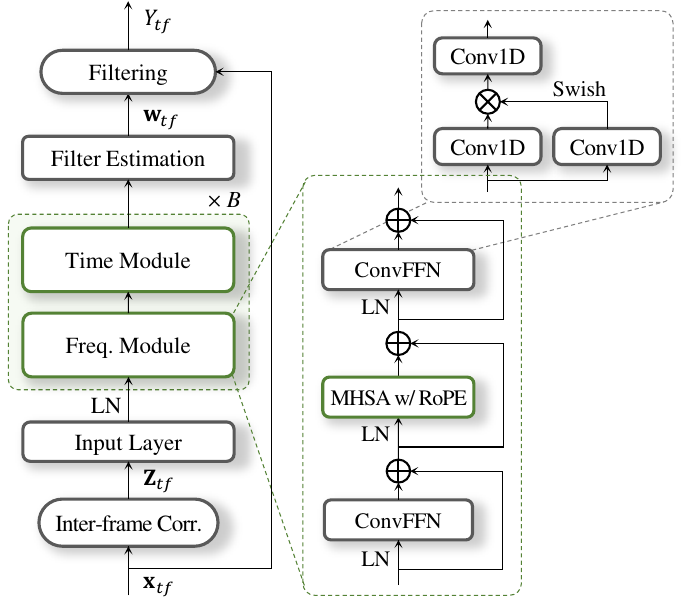}
    \vspace{-1mm}
    \caption{Overall architecture of IF-CorrNet. Inter-frame correlations are processed by the input layer and the frequency and time modules to estimate multi-frame filters.}
    \vspace{-1mm}
    \label{fig:architecture}
\end{figure}

\subsection{Time-frequency module}
The inter-frame correlations $\mathbf{Z}_{tf}$ are flattened, and their real and imaginary components are concatenated to form an input tensor of shape
$\mathbb{R}^{2(2L\hspace{-.3mm}+\hspace{-.3mm}1)^2\hspace{-.3mm}\times\hspace{-.3mm}T\hspace{-.3mm}\times\hspace{-.3mm}F}$.
As shown in Figure~\ref{fig:architecture}, the input layer projects the correlations into
$\mathbb{R}^{C\hspace{-.3mm}\times\hspace{-.3mm}T\hspace{-.3mm}\times\hspace{-.3mm}F}$ using two Conv2D layers with kernel sizes $(1,1)$ and $(3,3)$ and output channels $2C$ and $C$, respectively.
SwiGLU~\cite{swiglu} is used between the convolution layers, followed by layer normalization (LN).
Unlike global LN~\cite{TF_GridNet_TASLP,TF_Locoformer}, LN is applied over feature channels independently at each time-frequency bin.

The resulting features are processed by $B$ stacked frequency and time modules based on a modified Transformer~\cite{TF_Locoformer}.
The frequency module models inter-frequency dependencies by treating the features as $T$ sequences of length $F$, while the time module integrates these local correlation features across $T$ frames independently for each frequency bin.
Finally, a linear layer projects the features to
$\mathbb{R}^{2(2L\hspace{-.3mm}+\hspace{-.3mm}1)\hspace{-.3mm}\times\hspace{-.3mm}T\hspace{-.3mm}\times\hspace{-.3mm}F}$, which is reshaped to estimate the complex multi-frame filter $\mathbf{w}_{tf}$.

\subsection{Transformer block with ConvFFN module}

Local time-frequency context is important for speech, and self-attention alone does not necessarily produce locally smooth attention patterns. Each frequency and time module therefore employs a macaron-style Transformer block~\cite{macaronnet} with two convolution-augmented feed-forward network (ConvFFN) modules, following TF-Locoformer~\cite{TF_Locoformer}.
As illustrated in Figure~\ref{fig:architecture}, the block consists of multi-head self-attention (MHSA) with rotary positional encoding (RoPE)~\cite{RoPE}, surrounded by two ConvFFN modules.
Each ConvFFN replaces the point-wise layers of a conventional feed-forward network with two 1D convolution layers to enhance local context modeling, with SwiGLU used as the hidden activation.

\section{Experimental Setup}

\subsection{Datasets and evaluation}
We conducted experiments on the REVERB Challenge corpus~\cite{REVERB}, which contains 16-kHz simulated and real recordings with speaker-disjoint training, development, and evaluation sets. The training set consists of 7,861 clean utterances from WSJCAM0~\cite{WSJCAM0}, from which reverberant mixtures were generated using 24 measured room impulse responses (RIRs) with reverberation times ($T_{60}$) ranging from 0.2 s to 0.8 s and diffuse background noise at a signal-to-noise ratio (SNR) of 20 dB. The training target $D$ is generated using only the RIR peak. 
The proposed models were trained only on these simulated mixtures without additional data augmentation~\cite{DCN}.
The development and evaluation sets contain simulated (``SimData'') and real (``RealData'') recordings. SimData contains 1,484 development and 2,176 evaluation utterances from WSJCAM0, generated using impulse responses measured in three rooms at near and far microphone distances with stationary background noise. RealData contains 179 development and 372 evaluation utterances recorded in an unseen physical room at source--microphone distances of 1 m and 2.5 m, providing a mismatched real-world evaluation condition.

The checkpoint was selected using the SimData development set and evaluated on both SimData and RealData. We report cepstral distance (CD), log-likelihood ratio (LLR), frequency-weighted segmental SNR (SNR$_{\mathrm{fw}}$), perceptual evaluation of speech quality (PESQ), and speech-to-reverberation modulation energy ratio (SRMR). Since reference clean signals are unavailable for RealData, CD, LLR, SNR$_{\mathrm{fw}}$, and PESQ were evaluated only on SimData, while SRMR was also evaluated on RealData.

\subsection{Training and model configuration}

IF-CorrNet and its smaller variant were trained with AdamW for up to 40 epochs using 4-second segments and a batch size of 2. STFT was computed with a 512-sample Hann window and a hop size of 256.
For IF-CorrNet, we set $B\hspace{-.5mm}=\hspace{-.5mm}6$, $C\hspace{-.5mm}=\hspace{-.5mm}96$, and the ConvFFN hidden dimension and kernel size to $C_H\hspace{-.5mm}=\hspace{-.5mm}192$ and $K\hspace{-.5mm}=\hspace{-.5mm}7$, respectively. We also trained a smaller variant with $B\hspace{-.5mm}=\hspace{-.5mm}6$, $C\hspace{-.5mm}=\hspace{-.5mm}64$, $C_H\hspace{-.5mm}=\hspace{-.5mm}128$, and $K\hspace{-.5mm}=\hspace{-.5mm}3$. The number of MHSA heads was set to 4.
The training objective combines time-domain $\ell_1$ loss and time-frequency-domain multi-resolution $\ell_1$ loss~\cite{Loss}, using FFT window sizes of $\{256, 512, 768, 1024\}$. The training protocol follows~\cite{DCN}, and all configurations compared in Tables~\ref{tab:ablation} and~\ref{tab:efficiency} are trained on the same simulated data under this protocol; models that directly estimate spectra use input normalization and inverse rescaling.

\begin{figure}
\centering
\subfloat[]{\includegraphics[width=0.238\textwidth]{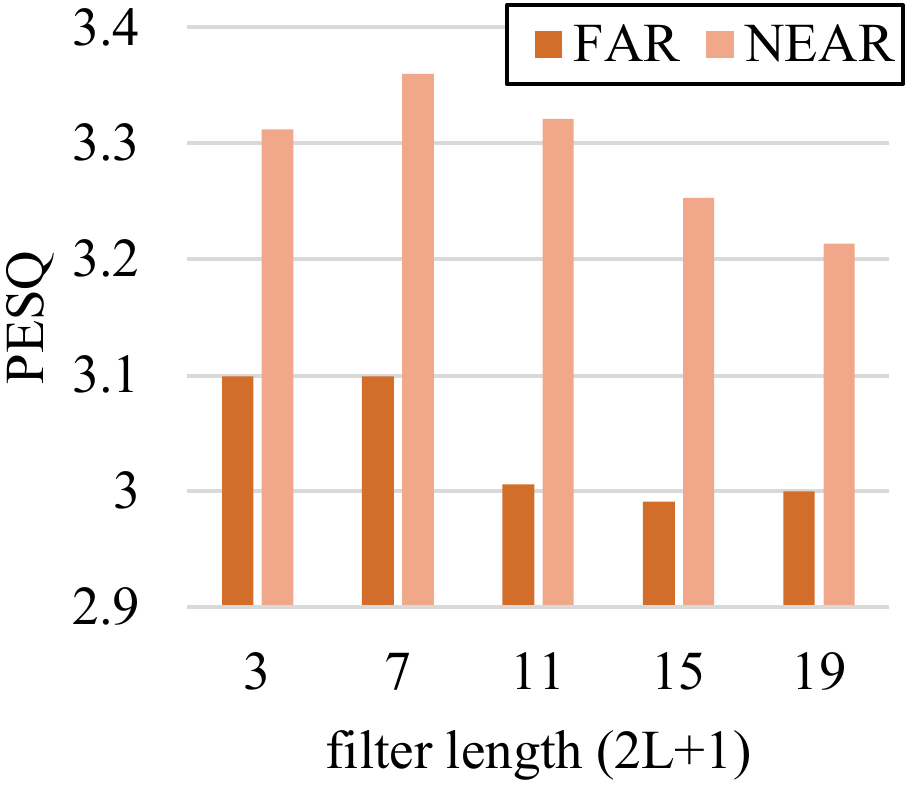}}\label{fig:PESQ}\hspace{-1mm}
\subfloat[]{\includegraphics[width=0.238\textwidth]{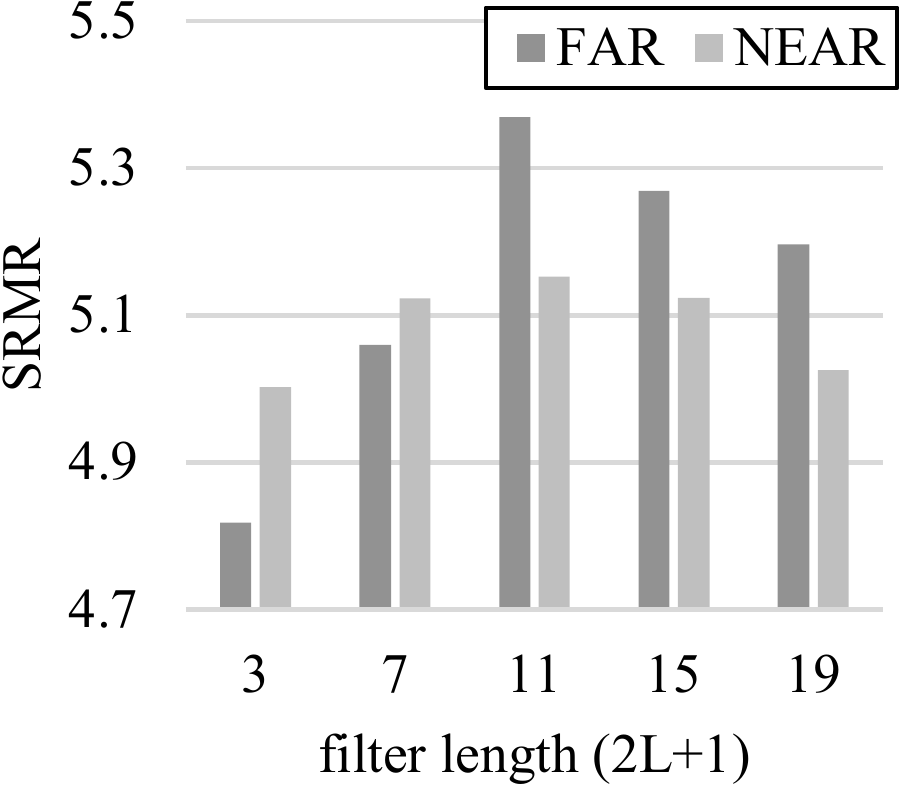}}\label{fig:SRMR}\hspace{0mm}
\vspace{-1mm}
\caption{SimData (a) PESQ and (b) SRMR versus filter length $2L+1$,
shown for near and far microphone conditions.}
\vspace{-1mm}
\label{fig:plot_taps}
\end{figure}

\vspace{1mm}

\section{Experimental Results}

\subsection{Effect of filter length}

Figure~\ref{fig:plot_taps} reports PESQ and SRMR on the SimData evaluation set for varying half-width $L$ under near- and far-microphone conditions.
Increasing $L$ jointly expands the correlation input window and output filter support.
PESQ peaks at $L=3$ (seven coefficients), whereas SRMR peaks around $L=5$ (eleven coefficients), indicating metric-dependent behavior.
Far-field SRMR eventually exceeds near-field SRMR as $L$ increases.
Considering both metrics and microphone distances, we use $L=3$ for the remaining experiments.

\begin{table}
\small
\caption{Performance comparison of existing methods on the REVERB Challenge evaluation set. Baseline results are taken from~\cite{DCN}.}
\vspace{-2mm}
\setlength{\tabcolsep}{1.1pt}
\def\arraystretch{1.03}
\begin{center}
\scalebox{0.96}{\begin{tabular}{lcccccc}
\toprule
\multirow{2}{*}{\hspace{-0.1mm}\textbf{Method}\hspace{-0.1mm}} & \multicolumn{5}{c}{\textbf{SimData}} & \textbf{RealData}\\
\cmidrule(lr){2-6}\cmidrule(lr){7-7}
& {\textbf{CD}}\textsuperscript{$\downarrow$} & {\textbf{SRMR}}\textsuperscript{$\uparrow$} & {\textbf{LLR}}\textsuperscript{$\downarrow$} & {\textbf{SNR}\textsubscript{fw}}\textsuperscript{$\uparrow$} & {\textbf{PESQ}}\textsuperscript{$\uparrow$} & \textbf{SRMR}\textsuperscript{$\uparrow$}\\
\midrule
No Processing& 3.975&3.687&0.574&\hspace{1.5mm}3.617&1.503&3.180\\
WPE~\cite{WPE_2012}&3.748&4.220&0.514&\hspace{1.5mm}4.864&1.722&3.978\\
GCRN~\cite{GCRN}&2.534&4.861&0.325&10.753&1.934&4.846\\
TCN+SA~\cite{TCN_SA}&2.200&5.170&0.240&13.060&2.580&5.540\\
D-MFMVDR~\cite{Tammen2021_MFMVDR}&2.639&4.892&0.316&\hspace{1.5mm}9.649&2.167&5.530\\
DCN~\cite{DCN}&{2.001}&\textbf{5.269}&0.225&13.326&2.935&6.476\\
\midrule
IF-CorrNet(\textit{small}) & 2.359 & 4.966 & 0.233 & 14.295 & 3.016 & 7.009\\
\rowcolor{Gray}
IF-CorrNet&\textbf{1.899}&{5.091}&\textbf{0.179}&\textbf{15.663}&\textbf{3.230}&\textbf{7.548}\\
\bottomrule
\end{tabular}}
\end{center}
\vspace{-1mm}
\label{tab:baselines}
\end{table} 

\subsection{Comparison with existing baselines}
\label{subsec:main}
We first compare IF-CorrNet with representative single-channel dereverberation baselines on the REVERB Challenge dataset: weighted prediction error (WPE)~\cite{WPE_2012}, GCRN~\cite{GCRN}, TCN+SA~\cite{TCN_SA}, D-MFMVDR~\cite{Tammen2021_MFMVDR}, and DCN~\cite{DCN}. Table~\ref{tab:baselines} summarizes results on the SimData and RealData evaluation sets.
Overall, IF-CorrNet achieves the best performance on SimData across CD, LLR, SNR$_{\mathrm{fw}}$, and PESQ, while also providing a substantial SRMR gain on RealData.
IF-CorrNet(\textit{small}) also achieves a RealData SRMR of 7.009, showing that the proposed formulation remains effective with reduced model capacity.

\begin{table}
\small
\caption{Input/output ablation on REVERB. SF/MF: single-/multi-frame.}
\vspace{-3mm}
\setlength{\tabcolsep}{1.5pt}
\def\arraystretch{1.05}
\begin{center}
\scalebox{0.96}{\begin{tabular}{cccccccc}
\toprule
\multirow{2}{*}{\textbf{Input}} & \multirow{2}{*}{\textbf{Output}}& \multicolumn{5}{c}{\textbf{SimData}} & \textbf{RealData}\\
\cmidrule(lr){3-7}\cmidrule(lr){8-8}
&& {\textbf{CD}}\textsuperscript{$\downarrow$} & {\textbf{S\hspace{-.2mm}R\hspace{-.2mm}M\hspace{-.2mm}R}}\textsuperscript{$\uparrow$} & {\textbf{LLR}}\textsuperscript{$\downarrow$} & {\textbf{SNR}\textsubscript{fw}}\textsuperscript{\hspace{-.5mm}$\uparrow$} & {\textbf{P\hspace{-.2mm}E\hspace{-.2mm}S\hspace{-.2mm}Q}}\textsuperscript{$\uparrow$} & \textbf{SRMR}\textsuperscript{$\uparrow$}\\
\midrule
\rowcolor{Gray}
IF-Corr. & MF-Filter &\textbf{1.899}&\textbf{5.091}&\textbf{0.179}&\textbf{15.663}&3.230 & \textbf{7.548}\\
IF-Corr. & SF-Mask &1.938 & 5.041 & 0.188 & 15.644 & 3.248 & 7.462\\
SF-Raw & MF-Filter &2.003&{5.022}&0.191&15.471& 3.246 & 7.225\\
MF-Raw & MF-Filter &2.399&5.059&0.229&14.782&\textbf{3.268} & 7.231\\
SF-Raw & SF-Mask &2.006&5.011&0.198&15.453&{3.244} & 7.245\\
SF-Raw & Mapping &2.173&5.058&\textbf{0.179}&15.241&3.207 & 7.091\\
\bottomrule
\end{tabular}}
\end{center}
\vspace{-3mm}
\label{tab:ablation}
\end{table} 

\subsection{Impact of inter-frame correlations on robustness}

Table~\ref{tab:ablation} examines the inter-frame correlation (IF-Corr.) input and multi-frame filtering (MF-Filter), alongside a single-frame raw STFT input (SF-Raw), single-frame complex masking (SF-Mask), and direct spectral mapping.
With SF-Mask, replacing SF-Raw with IF-Corr. improves all reported
SimData metrics and raises RealData SRMR from 7.245 to 7.462.
The correlation representation thus benefits enhancement even without
multi-frame output filtering.
With MF-Filter, it raises RealData SRMR from 7.225 to 7.548.
MF-Filter improves RealData SRMR over SF-Mask when paired with IF-Corr.,
whereas the two outputs perform similarly with SF-Raw, indicating that
the multi-frame output benefits from explicit correlation inputs.
Among the IF-Corr. configurations, MF-Filter yields better SimData CD,
SRMR, LLR, and SNR$_{\mathrm{fw}}$, while SF-Mask yields higher PESQ
(3.248 versus 3.230).

Stacking seven raw STFT frames (MF-Raw) gives a RealData SRMR similar to SF-Raw (7.231 versus 7.225). This is consistent with the encoder convolutions and time modules already providing temporal context for SF-Raw, suggesting that explicit inter-frame correlations are more useful than frame stacking alone.
With direct spectral mapping, SF-Raw reaches a SimData PESQ of 3.207 and a RealData SRMR of 7.091, close to the SF-Raw filtering and masking variants (7.225 and 7.245). Overall, replacing SF-Raw with IF-Corr. produces a larger RealData SRMR gain than changing the output formulation.

\subsection{Efficiency and distance robustness analysis}

We further compare IF-CorrNet with recent enhancement baselines, including TF-GridNet~\cite{TF_GridNet_TASLP}, MP-SENet~\cite{mpsenet}, and TF-Locoformer~\cite{TF_Locoformer}, trained under the same REVERB protocol. Table~\ref{tab:efficiency} reports performance, model complexity, and runtime.

IF-CorrNet achieves the highest enhancement scores among the systems in Table~\ref{tab:efficiency}.
On an RTX 5090 GPU, its real-time factor (RTF) is 0.025 versus 0.081 for TF-GridNet, despite a higher multiply-accumulate (MAC) cost (161.4 versus 127.3 GMAC/s).
IF-CorrNet(\textit{small}) has an RTF of 0.024, close to MP-SENet's 0.022, while retaining competitive performance.
These baselines also differ in how their SimData performance transfers to RealData: TF-GridNet keeps a high RealData SRMR (7.319), whereas TF-Locoformer, with a SimData SNR$_{\mathrm{fw}}$ between those of the other baselines, yields the lowest RealData SRMR (6.639). Audio examples are available on the demo page.

Table~\ref{tab:distance} compares near- and far-field performance on SimData.
IF-CorrNet achieves the highest far-field scores and the smallest PESQ gap (0.261), while its SNR$_{\mathrm{fw}}$ gap (2.240) is slightly larger than DCN's (2.091).

\begin{table}
\small
\caption{Performance and computational cost of recent baselines under the same REVERB training protocol. RTF is measured on an NVIDIA GeForce RTX 5090.}
\vspace{-3mm}
\setlength{\tabcolsep}{1.5pt}
\def\arraystretch{1.02}
\begin{center}
\scalebox{0.98}{\begin{tabular}{lcccccc}
\toprule
\multirow{2}{*}{\hspace{-0.1mm}\textbf{Method}\hspace{-0.1mm}} & \textbf{Param} & \textbf{MAC} & \multirow{2}{*}{\textbf{RTF}} & \multicolumn{2}{c}{\textbf{SimData}} & \textbf{\hspace{-.5mm}RealData}\\[-2pt]
\cmidrule(lr){5-6}\cmidrule(lr){7-7}
& (M) & (G/s) & & {\textbf{SNR}\textsubscript{fw}}\textsuperscript{\hspace{-.5mm}$\uparrow$} & {\textbf{PESQ}}\textsuperscript{$\uparrow$} & \textbf{SRMR}\textsuperscript{$\uparrow$}\\
\midrule
No Processing& - & - & - & 3.617&1.503&3.180\\
\midrule
TF-GridNet & 8.4 & 127.3 & 0.081 & 14.458 & 2.825 & 7.319 \\
MP-SENet & 2.3 & 22.2 & 0.022 & 13.622 & 2.811 & 7.007 \\
TF-Locoformer & 15.0 & 258.2 & 0.028 & 14.102 & 2.642 & 6.639\\
\midrule
IF-CorrNet(\textit{small}) & 2.1 & 33.7 & 0.024 & 14.295 & 3.016 & 7.009\\
\rowcolor{Gray}
IF-CorrNet& 10.0 & 161.4 & 0.025 &\textbf{15.663}&\textbf{3.230}&\textbf{7.548}\\
\bottomrule
\end{tabular}}
\end{center}
\label{tab:efficiency}
\vspace{-3mm}
\end{table} 

\begin{table}
\small
\caption{SimData distance comparison ($\Delta=\mathrm{Near}-\mathrm{Far}$; SNR$_{\mathrm{fw}}$ in dB). Bold indicates best among processed systems.}
\vspace{-3mm}
\setlength{\tabcolsep}{2.4pt}
\def\arraystretch{1.01}
\begin{center}
\scalebox{0.98}{\begin{tabular}{lcccccc} %
\toprule
\multirow{2}{*}{\textbf{Method}} & \multicolumn{3}{c}{\textbf{PESQ}} & \multicolumn{3}{c}{\textbf{SNR}\textsubscript{fw}} \\
\cmidrule(lr){2-4}\cmidrule(lr){5-7}
& \textbf{Far} & \textbf{Near} & \textbf{Diff.(\hspace{-.3mm}$\Delta$\hspace{-.3mm})} & \textbf{Far} & \textbf{Near} & \textbf{Diff.(\hspace{-.3mm}$\Delta$\hspace{-.3mm})} \\
\midrule
No Processing & 1.339 & 1.667 & 0.328 & \hspace{1.5mm}2.650 & \hspace{1.5mm}4.582 & 1.932 \\
\midrule
DCN\cite{DCN} & 2.553 & 3.317 & 0.764 & 12.280 & 14.371 & \textbf{2.091} \\
TF-GridNet & 2.503 & 3.149 & 0.646 & 13.003 & 15.913 & 2.910 \\
MP-SENet & 2.566 & 3.056 & 0.490 & 12.414 & 14.829 & 2.415 \\
TF-Locoformer & 2.385 & 2.900 & 0.515 & 12.542 & 15.663 & 3.121 \\
\midrule
IF-CorrNet(\textit{small}) & 2.861 & 3.170 & 0.309 & 13.020 & 15.570 & 2.550 \\
\rowcolor{Gray}
IF-CorrNet & \textbf{3.099} & \textbf{3.360} & \textbf{0.261} & \textbf{14.543} & \textbf{16.783} & {2.240} \\
\bottomrule
\end{tabular}}
\end{center}
\label{tab:distance}
\vspace{-3mm}
\end{table}

\section{Conclusion}
\vspace{-1mm}
We introduced IF-CorrNet, a deep filter estimation model for monaural speech dereverberation that explicitly leverages inter-frame correlations rather than raw complex STFT inputs. By exploiting correlations between adjacent time frames as the primary network input, IF-CorrNet directly estimates multi-frame dereverberation filters.
Experiments on REVERB show improvements over the compared baselines on CD, LLR, SNR$_{\mathrm{fw}}$, and PESQ for SimData, together with higher SRMR on RealData. The comparisons further show that performance on simulated data does not transfer equally to real recordings across architectures, while IF-CorrNet outperforms the recent enhancement baselines on both SimData and RealData.
Future work includes multi-channel extensions with spatio-temporal correlations and joint denoising, dereverberation, and separation.

\vfill\pagebreak

\bibliographystyle{IEEEbib_compact}
\small
\bibliography{refs}

\end{document}